\documentstyle[prl,aps]{revtex}

\begin{document}

\title{\bf On the Local Finite Density Relativistic Quantum Field Theories}

\draft

\author{S. Ying}
\address{Physics Department, Fudan University,
Shanghai 200433, People's Republic of China}
\def\bra#1{\mathopen{\langle#1\,|}}
\def\ket#1{\mathclose{|\,#1\rangle}}

\maketitle
\begin{abstract}
   It is found that, in addition to the conventional ones, a local
approach to the relativistic quantum field theories at both zero and
finite density consistent with the violation of Bell like inequalities
should contain, and provide solutions to at least three additional
problems, namely, 1) the statistical gauge invariance 2) the dark
components of the local observables and 3) the fermion statistical
blocking effects, base upon an asymptotic non-thermo ensemble. An
application to models are presented to show the importance of the
discussions.
\end{abstract}


Besides its relevance to domestic processes like in heavy ion
collisions, in nuclear matter, a consistent theory for relativistic
finite density systems is neither less interesting fundamentally due
to its relevance to profound cosmological questions like the mechanism
for baryogenesis, the nature of dark matter, etc., nor can it be
expected to be trivially derivable from the theoretical framework at
zero density.  Although such kind of theory, which adopt the basic
framework of the field theoretical representation of non-relativistic
many body systems exists and is widely used in literature, one
essential ingredient of the relativistic spacetime, namely the
principle of locality, has not been properly addressed. This is
because in the grand canonical ensemble (GCE) base on which such a
theory is built, the chemical potential is a global quantity in
spacetime.  As it is well know in the studies of few and many body
relativistic systems that it is not enough to simplly adopt the
relativistic kinematics for each particles in the system when the
interaction is present.  The less well understood relative time
between particles inside the system exists due to the lack of a frame
independent definition of simultaneity in the relativistic spacetime
which constitutes another foundamental difference between
non-relativistic systems based upon Newtonian spacetime and the
relativistic ones.

A measurement of the ground state fermion number density in GCE
given by
\begin{eqnarray}
\overline n &=& \lim_{T\to 0,\Omega\to\infty }
         {T\over\Omega} {\partial \ln Z\over \partial \mu}
\label{gl-rho}
\end{eqnarray}
corresponds to a global measurement. Here $Z$ is the partition
function, $T$ is the temperature, $\mu$ is the chemical potential and
$\Omega$ is the volume of the system. Global measurements are not
directly definable since simultaneity of two evens separated from each
other by a space-like distance depends on reference frame in
relativity. In addition, the property of asymptotic freedom in strong
interaction makes the local measurements relevant to the study of the
properties of fundamental current quarks at short distances where they
are point like and almost free particles.

  Local measurements in quantum field theory are realized by
exerting an external local field to the system at the spacetime point
interested, and, the measured quantity are deduced from the response
of the system to the external field. For finite density systems, a
Lorentz 4-vector local field $\mu^\alpha(x)$, called the primary
statistical gauge field, is introduced. 
The ground state expectation value of fermion number density can be
expressed as the functional derivative
\begin{eqnarray}
    \overline \rho(x) &=& {\delta \ln Z\over \delta\mu^0(x)}
\label{lc-rho}
\end{eqnarray}
with $Z$ a functional of $\mu^\alpha$.
It corresponds to a local measurement in the ground state at the
spacetime point x.

That $\overline\rho\equiv \overline n$ is not mathematically
warranted. Because $\overline \rho$ contains not only the
contributions from the quasiparticles but also the ones from localized
transient fluctuations of the fields \cite{dark} while $\overline n$
contains only the quasiparticle contributions.  The difference $\rho_D
= \overline\rho - \overline n$ is called the dark component of the
local observable.

$\overline\rho $ follows from the asymptotic grand canonical ensemble
(AGCE), which is defined as the GCE of the corresponding free system
in the remote past (or future) when the interaction effects are
adiabatically switched off, with the Lagrangian density that underlies
$Z$ given by
\begin{eqnarray}
    {\cal L}' &=& {1\over 2}\overline\Psi (i\rlap\slash\partial +
      \rlap\slash\hspace{-1pt}\mu O_3 - m)
     \Psi + {\cal L}_B + {\cal L}_{\mbox{\scriptsize int}}
\end{eqnarray}
and $m$ the fermion mass, ${\cal L}_B$ the Lagrangian density of boson
fields of the system and ${\cal L}_{\mbox{\scriptsize int}}$ the
interaction between the boson and fermion fields. An 8 component
``real'' fermion field $\Psi$ is used for the discussion
\cite{8comp-pap} with $O_3$ the third Pauli matrices acting on the
upper and lower 4-component of $\Psi$. In most of the cases, ${\cal
L}_{\mbox{\scriptsize int}}=-\overline\Psi(\Sigma(f)-m)\Psi$ with
$\Sigma$ a function(al) of the boson fields represented by $f$ so that
the fermion degrees of freedom can be integrated out leading to an
effective Euclidean action for the boson fields
\begin{eqnarray}
          S_{\mbox{\scriptsize eff}}[f,\mu] &=& {1\over 2} \mbox{SpLn} {
          [i\rlap\slash\partial +
          \rlap\slash\hspace{-1pt}\mu O_3-\Sigma(f)]\over 
          [i\rlap\slash\partial +
          \rlap\slash\hspace{-1pt}\mu O_3-m]}
           + \int d^4 x \left [{\cal L
          }_B(f) + \mu\overline\rho-\overline e 
          \right ]
\label{eff-action}
\end{eqnarray}
with ``Sp'' the functional trace, $\overline e$ the average energy
density of the corresponding free system of fermions ($\Sigma=0$) and
$\overline\rho$ is the local average of fermion number density to be
discussed in the following at the ``chemical potential''
$\mu\equiv\sqrt{\overline\mu^\alpha \overline\mu_\alpha}$, where
$\overline \mu^\alpha$ is the ground state value of $\mu^\alpha$
normally provided by the external conditions.  The partition
functional $Z=\int D[f] exp(S_{\mbox{\scriptsize eff}})$ is then a
functional of $\mu^\alpha$ and $\ln Z$ its effective action. 

The first term in Eq. \ref{eff-action} is invariant under the
following $U(1)$ statistical gauge transformation $\mu^\alpha(x)\to
\mu^\alpha(x)-\partial^\alpha\Lambda(x)$, which corresponds to the
conservation of the fermion number.  Since $\mu^\alpha$ is a local
field, its excitation represents certain collective excitations of the
system.  Such a view introduces infinite many extra degrees of freedom
since there is no such a field in the original
theory. Superselection-sector in the representing Hilbert space
(SIRHS) containing physical states correspond to the primary
statistical gauge field exists and can be selected using a ``Gauss
Law'' constraint. For the statistical gauge invariant system, it can
be imposed differently on the physical eigenstates of the Hamiltonian
without causing contradictions. States in a SIRHS identified by a
coordinate dependent complex function $\varsigma$ satisfy
\begin{eqnarray}
     \bra{\psi^i_\varsigma}
       \left (\widehat \rho + \nabla \cdot \widehat\pi_{u}  \right )
      \ket{\psi^j_\varsigma}
      &=& \delta_{E_i E_j} N_{ij} \varsigma
\label{Q-eigenstate}
\end{eqnarray}
with $\widehat\pi_{u}$ the ``statistical electric field'',
$\ket{\psi^k_\varsigma}$ a physical state that has energy $E_k$,
$\delta_{EE'}$ taking zero or unity value if $E\ne E'$ or $E'=E$
(assuming that $E$ is discrete before thermodynamic limit is taken)
and $N_{ij}$ independent of spacetime. $\widehat
\rho + \nabla \cdot \widehat\pi_{u}$ is the generator of local
statistical gauge transformations, so the physical states in a SIRHS
change a common (coordinate dependent) phase under a specific gauge
transformation rather than remains invariant, which has hitherto been
used. Clearly, the later condition is a special case for the former.

Due to the conservation of the fermion
number, the SIRHS of the system with fixed fermion number selected
\cite{Huang} by the AGCE is invariant during the time evolution. In
the AGCE, the dependence of $\overline\rho$ upon the spacetime
independent $\mu$ remains the same whether there is interaction in the
system or not.  It is given by $\overline\rho = N_g \mu^3/3\pi^2$ for
a massless system, with $N_g$ the total internal degrees of freedom
besides the spin of the fermion. $\overline n$ does not has a simple
dependence on $\mu$, especially when there is certain kind of phase
transition that modifies the excitation spectra of the system.

In some sense, the AGCE is a GCE for the initial state of the system
under the time evolution, it is a canonical ensemble for the
interaction effects since the invariant SIRHS (under the time
evolution) is fixed in the remote past (or future).  The usefulness of
the AGCE in numerical simulations has already manifested in
Monte-Carlo studies of the finite density problems \cite{Kogut} based
upon GCE for a finite lattice.  It is shown using the chiral
Gross-Neveu model that for a given chemical potential, the fermion
number density of the system has two stable values, which are reached
by generating the Monte-Carlo ensemble at either finite or zero
density. The first stable one correspond to the one given by the AGCE
$\rho-\mu$ relation and the second one correspond to the GCE
prediction for a massive system of quasi-particles.  The exact results
for the global treatment of the problem crossover from the later
$\rho-\mu$ curve to the former one near the chiral symmetry
restoration $\mu=\mu_c$. This behavior can be interpreted as that
before the chiral symmetry restoration, the first $\rho-\mu$ curve for
the AGCE is a local minimum of the (Euclidean) action of the various
configurations in the Monte-Carlo ensemble with the GCE curve for the
massive system of quasiparticles the absolute one. Near the chiral
symmetry restoration point, their role exchanges. The existence of a
stable local minimum for the AGCE $\rho-\mu$ curve constitutes a
strong numerical evidence that the AGCE is useful in numerical
simulations of the localized theories. In fact, it is expected that
some of the puzzles encountered in the lattice gauge theory studies at
finite density, like the early on set of the baryon number before the
chiral symmetry restoration, can be understood in the light of the
AGCE since the finite size effects of the lattice gauge theory study
based upon GCE corresponds to, to some degree, a low resolution local
measurement of an infinite system \cite{dark,8comp-pap}.

This interpretation can be substantiated by using a cluster
decomposition approximation of the partition functional $Z[\mu]$
\cite{dark} which is expected to provide a non-perturbative picture
for the $\mu$ dependence of the fermion number density at sufficiently
low spacetime resolutions. In the crudest approximation, the ground
state fermion number density can be written as \cite{dark}
\begin{eqnarray}
\rho_{\Delta\Omega}(\overline\sigma,\mu)  &=& \sqrt{\alpha\over\pi}
       \int_{-\mu-\overline\sigma}^{\mu-\overline\sigma} d\sigma'
        e^{-\alpha{\sigma'}^2} \rho_\infty(\overline\sigma+\sigma',\mu),
\end{eqnarray}
where $\bar\sigma$ is the ground state expectation value of $\sigma$,
$\rho_\infty(\sigma,\mu)= g (\mu^2-\sigma^2)^{3/2}/3\pi^2$ is the
ground state fermion number density in the GCE with fermion mass
$\sigma$. Here $\Delta\Omega$ is the spacetime volume that can be
resolved in the observation.  $\alpha$ is proportional to
$\Delta\Omega$ for large enough $\Delta\Omega$ and approaches a
constant value for small $\Delta\Omega$ which scales with the inverse
mass gap of the $\sigma'$ excitation of the system. Clearly,
$\rho_{\Delta\Omega}(\overline\sigma,\mu)\to \rho_\infty
(\overline\sigma,\mu) = \overline n$ as $\Delta\Omega\to\infty$.  This
qualitative picture shows that the result of a local measurement of
the fermion number density approaches that of the global one as the
resolution gets lowered; in the limit of the zero resolution, the
result approaches that of the global measurement as expected.

The existence of the dark component of local fermion number density
has interesting implications. Take the vacuum state of a massless
strongly interacting system in which the chiral symmetry is
spontaneously broken down for example. The originally gapless vacuum
state acquires an excitation gap. If only the massive quasi-particles
are taken into account, the baryon number density are expected to be
non-vanishing only when $\mu$ is larger than the mass of the
quasi-particles. The local fermion number density is however finite as
long as $\mu$ is not zero. A natural question arises as to what this
dark component of fermion number corresponding to?  It is reasonable
to conjecture that this dark component of fermion number corresponds
to those fermion states that are localized and non-propagating
\cite{Stern} similar to the Anderson localization in a condensed
matter system. The random potential here is self-generated within the
system by the transient and localized random quantum fluctuations of
the $\sigma$ and other boson fields. Such a conjecture is amenable to
future studies.

At the fundamental level, the dark component of local observables is
of pure quantum in origin related to the space-like correlations
between local measurements in the relativistic quantum world that are
shown to exist in experimental observations \cite{Aspect,Tapster,Tittel},
which are what reality manifests itself \cite{local}. This is because
the size of the correlated cluster used in studying the vacuum state
(in the Euclidean spacetime) tends to zero in the classical limit
($\hbar\to 0$) as a result of the classical relativistic causality,
which suppresses the dark component for any finite resolution
observation \cite{dark}.

The Euclidean effective action for the boson field given by
Eq. \ref{eff-action} can be evaluated in the usual way
\cite{8comp-pap}. Since the effective action given by
Eq. \ref{eff-action} is a canonical functional of $\mu^\alpha$, we can
make a Legendre transformation of it, namely $\widetilde
S_{\mbox{\scriptsize eff}} = S_{\mbox{\scriptsize eff}} -\int d^4 x
\overline \mu_\alpha \overline j^\alpha$ with $\overline j^\alpha$ the ground
state current of fermion number density, to make it a canonical
functional of the fermion density in order to study the the stability
of the vacuum state against fluctuations in fermion number. 

The vacuum state is defined as the ground state with lowest possible
energy density amongst all other ground states which due to external
influences are in states with higher energy densities.

  The effective potential can be defined using $\widetilde
S_{\mbox{\scriptsize eff}}$ for spacetime independent background $f$
fields: $V_{\mbox{\scriptsize eff}}=-\widetilde S_{\mbox{\scriptsize
eff}}/V_4$ with $V_4$ the volume of the spacetime box that contains
the system.  The stability of the vacuum state against the
fluctuations in fermion number density around $\overline\rho=0 $ can
be studied using $V_{\mbox{\scriptsize eff}}$, which is a canonical
function of $\overline\rho$. Minimization of $V_{\mbox{\scriptsize
eff}} $ with respect to $\mu$ gives the vacuum value $\rho_{vac}$ of
the system since the vacuum $\overline\rho$ is a known function of the
vacuum $\mu$ in AGCE.

The local finite density theory based upon the AGCE developed after
considering the above ingredients is examined by applying it to the
chiral symmetry breaking phase of the half bosonized Nambu
Jona-Lasinio model for the 3+1 dimension and the chiral Gross-Neveu
model for 2+1 dimensions with its effective potential given by
\begin{eqnarray}
V_{\mbox{\scriptsize eff}} &=& - N_g \int {d^Dp\over (2\pi)^D}\left [  \ln
    \left ( 1+{\sigma^2 \over p_+^2} \right )    + \ln
    \left ( 1+{\sigma^2 \over p_-^2} \right ) \right ] \nonumber\\
   &&+ {1\over 4 G_0}
    \sigma^2 + \overline e,
\label{Veff1}
\end{eqnarray}
where ``$D$'' is the spacetime dimension, $G_0$ is the coupling
constant and $p^2_\pm = (p_0\pm i\mu)^2+\mbox{\boldmath{p}}^2$.  It
can be shown that $\partial^2 V_{\mbox{\scriptsize eff}}/\partial\mu^2
|_{\mu=0} < 0$ as long as $\sigma\ne 0$ and $D\ge 2$, which implies
that the $\sigma\ne 0$ state with $\overline\rho=0$ is not stable
against fermion number fluctuations. Such a conclusion is both
theoretically and physically unacceptable.

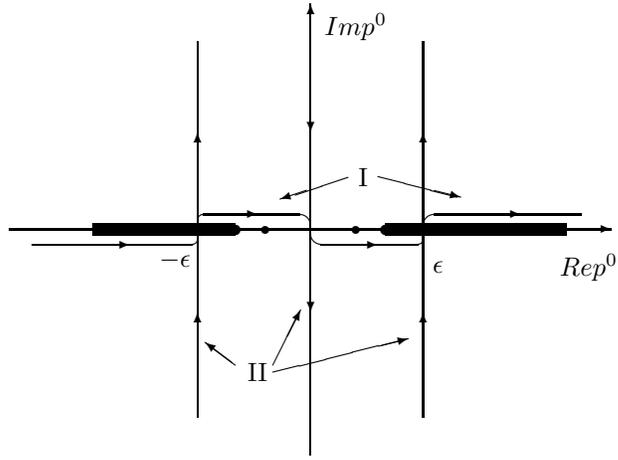
\begin{figure}
\unitlength=1.0mm
\linethickness{0.4pt}
\begin{picture}(120.00,110.00)
\put(40.00,80.00){\vector(1,0){80.00}}
\put(117.00,75.00){\makebox(0,0)[cc]{$Rep^0$}}
\put(86.00,107.00){\makebox(0,0)[cc]{$Imp^0$}}
\put(90.00,79.25){\rule{24.00\unitlength}{1.50\unitlength}}
\put(51.00,79.25){\rule{19.00\unitlength}{1.50\unitlength}}
\put(90.00,80.00){\circle*{1.50}}
\put(70.00,80.00){\circle*{1.50}}
\put(86.00,80.00){\circle*{1.00}}
\put(74.00,80.00){\circle*{1.00}}
\put(43.00,78.00){\line(1,0){20.00}}
\put(97.00,82.00){\line(1,0){19.00}}
\put(67.00,82.00){\line(1,0){11.00}}
\put(82.00,78.00){\line(1,0){11.00}}
\put(63.00,79.00){\oval(4.00,2.00)[rb]}
\put(67.00,81.00){\oval(4.00,2.00)[lt]}
\put(93.00,79.00){\oval(4.00,2.00)[rb]}
\put(97.50,80.50){\oval(5.00,3.00)[lt]}
\put(76.50,80.00){\oval(7.00,4.00)[rt]}
\put(82.00,81.00){\oval(4.00,6.00)[lb]}
\put(53.00,78.00){\vector(1,0){3.00}}
\put(70.00,82.00){\vector(1,0){3.00}}
\put(85.00,78.00){\vector(1,0){4.00}}
\put(102.00,82.00){\vector(1,0){6.00}}
\put(62.00,76.00){\makebox(0,0)[cc]{$-\epsilon$}}
\put(97.00,75.00){\makebox(0,0)[cc]{$\epsilon$}}
\put(80.00,50.00){\vector(0,1){60.00}}
\put(65.00,55.00){\line(0,1){50.00}}
\put(95.00,55.00){\line(0,1){50.00}}
\put(95.00,90.00){\vector(0,1){3.00}}
\put(65.00,90.00){\vector(0,1){3.00}}
\put(95.00,65.00){\vector(0,1){4.00}}
\put(65.00,65.00){\vector(0,1){4.00}}
\put(80.00,95.00){\vector(0,-1){2.00}}
\put(80.00,71.00){\vector(0,-1){2.00}}
\put(73.00,61.00){\makebox(0,0)[cc]{II}}
\put(70.00,62.00){\vector(-4,3){4.00}}
\put(75.00,61.00){\vector(4,1){18.00}}
\put(87.00,87.00){\makebox(0,0)[cc]{I}}
\put(85.00,87.00){\vector(-3,-1){9.00}}
\put(89.00,87.00){\vector(4,-1){11.00}}
\put(75.00,62.00){\vector(1,2){3.67}}
\end{picture}
\caption{\label{fig:contour} 
  The $p^0$ integration contour ${\cal C}$ for the effective
  potential. The thick lines extending to positive and negative
  infinity represent the branch cuts of the logarithmic function.
  Curve ``I'' is the original $p_0$ integration contour for the theory
  in the Minkowski spacetime. Curve ``II'' corresponds to the $p_0$
  integration contour for the theory in the Euclidean spacetime.}
\end{figure}

It appears that the AGCE is not sufficient for the local finite
density theory, a more general ensemble, called the asymptotic
non-thermo ensemble (ANTE) is introduced to cure this pathology. The
ANTE in the remote past of the system is not necessarily a strict
thermal one. This is because the $\sigma\ne 0$ phase, which is called
the $\alpha$-phase of the massless quark system, is known to be
condensed with macroscopic number of bare fermion-antifermion
pairs. These pairs occupy the low energy states of the system; they
block other fermions from further filling of these states. This kind
of statistical blocking effect is not encoded into Eq. \ref{Veff1},
which starts the time evolution of the system from a set of initial
states having zero number of fermion-antifermion pairs that do not
overlap with the states having a finite {\em density} of such pairs in
the thermodynamic limit. Its effects can however be included in the
boundary condition for the system like what has been done for the
finite density cases. Therefore, it is proposed that the initial
ensemble of quantum states in the ANTE for the system at zero
temperature and density are those ones with negative energy states
filled up to $E=-\epsilon$ rather than $E=0$ and with positive energy
states filled up to $E=\epsilon$ rather than empty. It is expected
that such a state can has sufficient overlap with the true vacuum
state of the system for properly determined $\epsilon$. In the ANTE
\cite{8comp-pap}, the effective potential corresponding to
Eq. \ref{Veff1} is expressed as
\begin{eqnarray}
V_{\mbox{\scriptsize eff}} &=& - N_g \int_{\cal C} 
   {d^Dp\over (2\pi)^D}\left [  \ln
    \left ( 1+{\sigma^2 \over p_+^2} \right )    + \ln
    \left ( 1+{\sigma^2 \over p_-^2} \right ) \right ] \nonumber\\
   &&+ {1\over 4 G_0}
    \sigma^2 + \overline e_{(+)} + \overline e_{(-)} - \overline e,
\label{Veff2}
\end{eqnarray}
where $\overline e_{(\pm)} = 2 N_g \int^{\mu_{\pm}} {d^{D-1}
p\over(2\pi)^{D-1}} p$.  Here the upper boundary of the radial $p$
integration in the D-1 dimensional momentum space is denoted as
$\mu_{\pm} \equiv \mu\pm \epsilon $. The contour ${\cal
C}$ for the (complex) $p_0$ integration is shown in
Fig. \ref{fig:contour} in which both the original Minkowski contour
and the Euclidean contour are drawn.  The effective potential for the
$\sigma\ne 0$ case in ANTE given by the above equation has two sets of
minima, the first set contains $\mu=\pm\mu_{vac}$ and $\epsilon=0$
points, the second one includes $\mu=0$ and $\epsilon= \pm
\epsilon_{vac}$ ones with finite $\mu_{vac}$ and $\epsilon_{vac}$. The
second solutions correspond to the absolute minima, namely, the vacuum
state. Although it is found that $ \epsilon_{vac} < \sigma_{vac}$ in
all cases that were studied, the effects of a finite vacuum $\epsilon$
can still have dynamical consequences, which is discussed in the
following and elsewhere \cite{8comp-pap,dark,pap2} due to the
existence of the dark component for local observables.

The vacuum and ground states of the strong interaction could have
different phases from the $\alpha$-phase
\cite{lettB,annPhy,Paps-a,Paps-b,Paps-c}. They are characterized by a
condensation of diquarks. Such a possibility is interesting because it
may be realized in the early universe, in astronomical objects and
events, in heavy ion collisions, inside nucleons \cite{Paps-b,nuc-pap}
and nuclei, etc.. For the possible scalar diquark condensation in the
vacuum, a half bosonized model Lagrangian is introduced
\cite{Model-I,8comp-pap}, which reads
\begin{eqnarray}
 {\cal L}_I & = & {1\over 2} \overline \Psi\left
             [i{\rlap\slash\partial}-\sigma- i\vec{\pi}\cdot
             \vec{\tau}\gamma^5 O_3-\gamma^5 {\cal A}_c\chi^c
             O_{(+)}-\gamma^5 {\cal A}^c\overline\chi_c O_{(-)} \right
             ]\Psi -{1\over 4 G_0} (\sigma^2 + \vec{\pi}^2) + {1\over
             2 G_{3'}} \overline\chi_c \chi^c ,\label{Model-L-1}
\end{eqnarray}
where $\sigma$, $\vec{\pi}$, $\overline\chi_c$ and $\chi^c$ are
auxiliary fields with $(\chi^c)^{\dagger} = - \overline\chi_c$ and
$G_0$, $G_{3'}$ are coupling constants of the model.  ${\cal A}_c$ and
${\cal A}^c$ $(c=1,2,3)$ act on the color space of the quark; they are
${\cal A}_{c_1c_2}^c = -\epsilon^{cc_1c_2}$ ${\cal A}_{c,c_1c_2} =
\epsilon^{cc_1c_2}$ with $\epsilon^{abc}$ ($a,b = 1,2,3$) the total
antisymmetric Levi--Civit\'a tensor. Here $O_{(\pm)}$ are raising and
lowering operators respectively in the upper and lowering 4 components
of $\Psi$.

This model has two non-trivial phases. The vacuum expectation of
$\sigma$ is non-vanishing with vanishing $\chi^2 \equiv-
\overline\chi_c \chi^c$ in the $\alpha$-phase. The vacuum state in the
$\alpha$-phase is condensed with quark-antiquark pairs. The vacuum
expectation of $\sigma$ is zero with finite $\chi^2$ that
spontaneously breaks the $U(1)$ statistical gauge symmetry in the
second phase, which is called the $\omega$-phase. There is a
condensation of correlated scalar diquarks and antidiquarks
in the color $\overline 3$ and $3$ states in the $\omega$-phase of the
vacuum state.

Since diquarks and antidiquarks are condensed in the
$\omega$-phase, it is expected that an exchange of the role of
$\epsilon$ and $\mu$ occurs. It is found to be indeed true: there
are also two sets of minima for the effective potential, the first set
is the one with finite $\mu=\pm\mu_{vac}$ and $\epsilon=0$ and the
second set corresponds to $\mu=0$ and finite
$\epsilon=\pm\epsilon_{vac}$ in the $\omega$-phase of the model. But
here the absolute minima of the system in the $\omega$-phase
correspond to the first set of solutions, in which the CP/T invariance
and baryon number conservation are spontaneously violated due to the
presence of a finite vacuum $\mu^\alpha_{vac}$. This conclusion is
also applicable to the $\beta$-phase of models \cite{lettB,annPhy}
with vector fermion pair and antifermion pair condensation, in which
the chiral symmetry $SU(2)_L\times SU(2)_R$ is also spontaneously
broken down.

The dynamics of the primary statistical gauge field are generated by
$S^{(\mu)}_{\mbox{\scriptsize eff}}[\mu]=\ln Z$. Due to the
statistical gauge invariance, the leading order in the derivative
expansion of $S^{(\mu)}_{\mbox{\scriptsize eff}}[\mu]$ in terms of
$\mu^\alpha$ can be expressed as the following: in the normal phase,
the effective action for slow varying
$\mu'_\alpha=\mu_\alpha-\overline\mu_\alpha $ is
\begin{eqnarray}
      S^{(\mu)}_{\mbox{\scriptsize eff}} 
       &=& \int d^4 x \left [ -{Z^{(\mu)}\over 4} f_{\mu\nu}
      f^{\mu\nu} +  {N_g\over \pi^2} {\overline \epsilon}^2 
       \mu' \cdot \mu' \right ] + \ldots 
\label{Seff-mu2}
\end{eqnarray}
with $f^{\alpha\beta} = \partial^\alpha
{\mu'}^\beta-\partial^\beta{\mu'}^\alpha$ and $\overline\epsilon$
the ground state value of $\epsilon$; in the phase where
local statistical $U(1)$ gauge symmetry is spontaneously broken down
and before considering electromagnetic interaction,
\begin{eqnarray}
      S^{(\mu)}_{\mbox{\scriptsize eff}} &=& {1\over 2}
         \int d^4 x \left [
             {i}\overline j^\alpha \overline j^\beta
       + g^{\alpha\beta} {\Pi^{(\mu)}}
       \right ] \mu'_\alpha
      \mu'_\beta + \ldots.
\label{Seff-mu3}
\end{eqnarray}
In both of the situations with slow varying $\mu'_\alpha$, $Z^{(\mu)}$
and $\Pi^{(\mu)}$, which can be extracted from the time-ordered
current--current correlator $< \mbox{T} j^\alpha(x) j^\beta(x')>$, are
approximately constants.

The excitation corresponds to $\mu^\alpha$ is massive for the vacuum
state in the $\alpha$-phase due to the statistical blocking effects.
There is no long range force associated with $\mu^\alpha$ in the
$\alpha$-phase.  It can be shown \cite{8comp-pap} that after including
the electromagnetic interaction, the spatial component of $\mu^\alpha$
excitation is massless for the vacuum state in the $\beta$- and
$\omega$- phases in the rest frame of matter.

In summary, it is found that general local relativistic quantum field
theory (in the sense of including both the zero and the finite density
situations) contains dark components for local observables,
statistical gauge invariance and ferminic blocking effects.  The
proper boundary condition for a consistent theory is that of an
optimal ANTE. This view seems to be receiving support from lattice
gauge theory studies \cite{sp1,sp2}. The rich implications
of the finding presented here remain to be explored.

\end{document}